\documentclass[aps,prl,reprint, amsmath, amssymb]{revtex4-1}
\usepackage{bm}
\usepackage[retainorgcmds]{IEEEtrantools}
\usepackage{graphicx}
\usepackage[caption=false]{subfig}
\usepackage{amsmath}
\usepackage{amsfonts}
\usepackage{amssymb}
\usepackage{color}
\usepackage{physics}
\usepackage{ragged2e}
\usepackage{tikz}
\usepackage{bbold}
\usepackage{centernot}

\usepackage{amstext} 
\usepackage{array}   
\usepackage{amsmath,tabu}
\usetikzlibrary{arrows,decorations,backgrounds,patterns}

\newcommand{\avg}[1]{\langle #1 \rangle}

\DeclareMathAlphabet{\mathcalligra}{T1}{calligra}{m}{n}

\usepackage{soul}

\usepackage{xcolor}
\usepackage[authormarkup=none]{changes}
\definechangesauthor[name={Eric}, color=red]{g}
\definechangesauthor[name={Franklin}, color=purple]{m}
\begin{document}

\title{Nonequilibrium thermodynamics of quantum coherence beyond linear response}

\author{Franklin L. S. Rodrigues}
\affiliation{Institute for Theoretical Physics I, University of Stuttgart, D-70550 Stuttgart, Germany.}
\author{Eric Lutz}
\affiliation{Institute for Theoretical Physics I, University of Stuttgart, D-70550 Stuttgart, Germany.}

\begin{abstract}
Quantum thermodynamics allows for the interconversion of  quantum coherence and mechanical work. Quantum coherence is thus a potential physical resource for quantum machines. However, formulating a general nonequilibrium thermodynamics of quantum coherence has turned out to be challenging.  In particular, precise conditions under which  coherence is beneficial to  or, on the contrary, detrimental for work extraction from a system have remained elusive. We here develop a generic dynamic-Bayesian-network approach to the far-from-equilibrium thermodynamics of coherence. We concretely derive generalized fluctuation relations and a maximum-work theorem that fully account for quantum coherence at all times, for both closed and open dynamics. We obtain criteria for successful coherence-to-work conversion, and identify a nonequilibrium regime where maximum work extraction is increased by quantum coherence for fast processes beyond linear response.


\end{abstract}
\maketitle{}

%
%

Coherence is a central feature of quantum theory. It is intimately associated with  linear  superpositions of states and related interference phenomena \cite{str17}. 
In the past decades, it has been recognized as an essential physical resource for quantum technologies  that can  outperform their classical counterparts \cite{nie00}, from quantum 
communication \cite{gis07} and quantum computation \cite{vin95} to quantum metrology \cite{gio11}. Understanding the role of quantum 
coherence in small-scale thermodynamics is a fundamental issue that has been examined using the formalism of open quantum systems 
\cite{scu03,scu11,har12,abe12,uzd15,kam16,cam19,fel03,pla14,kar16,bra16,bra17,bra20,rod19,san19,fra19} and the framework of  resource theory 
\cite{hor13,abe14,los15,nar15,los15a,cwi15,gou15,kor16,gou16,gou18,kwo18,los19,lob21}.  An important insight from these studies   is that  the laws of thermodynamics have to be 
extended to allow for the interconversion of coherence and energy \cite{los15,nar15,los15a,cwi15,gou15,kor16,gou16}. In particular, the  description of thermodynamic processes 
at low temperatures requires a generalization of  the usual  free energy in order to account for  coherent superpositions of energy eigenstates  of a system \cite{los15}.

The interplay between quantum mechanics and nonequilibrium thermodynamics is nontrivial, however. The crucial question under what conditions quantum coherence is also a useful physical resource in quantum thermodynamics has found no definite answer so far. 
 Depending on the considered problem, quantum coherence has indeed been theoretically predicted to  either enhance 
\cite{scu03,scu11,har12,abe12,uzd15,kam16,cam19} or decrease \cite{fel03,pla14,kar16,bra16,bra17,bra20} the amount of extractable work. Recent experimental realizations of
quantum heat engines are equally inconclusive, with one example reporting a performance boost  due to quantum coherence \cite{kla19}, and an other one observing an efficiency 
reduction linked to coherence-induced quantum friction \cite{pet19}.

We here develop a general nonequilibrium thermodynamics of quantum coherence valid arbitrarily far from equilibrium. We employ these findings  to clarify the impact of coherence  on quantum work 
extraction, and derive concrete criteria for successful coherence-to-work conversion, for both closed and open quantum dynamics. To this end, we derive  novel detailed and integral fluctuation relations for the nonequilibrium entropy production that fully account for  superpositions of energy 
levels at all times, especially at the beginning of a quantum process. Fluctuation theorems are fundamental  extensions of the second law   for small systems subjected to 
classical \cite{jar11,sei12} and quantum \cite{esp09,cam11} fluctuations. Their generic validity beyond the linear response regime makes them invaluable in the investigation of 
nonequilibrium phenomena. We concretely use  a powerful dynamic Bayesian network approach  that specifies the local dynamics of a system conditioned on the global evolution \cite{nea03,dar09}. As a result,  this formalism preserves the quantum properties  of the system at all times \cite{mic20,par20,str20,mic21}. This is at variance with other commonly applied  methods, such as the two-point-measurement scheme 
\cite{tal07}, that is not able to quantitatively capture  initial and final quantum coherences, since these are destroyed by local projective measurements  \cite{esp09,cam11}. We  furthermore obtain a quantum generalization of the 
maximum-work theorem that provides an upper bound to the amount of work that can be extracted from a system \cite{cal85}. The latter inequality includes a variation of the quantum coherence in the energy representation, expressed in terms of the relative 
entropy of coherence  \cite{kla19}. This result  depends  on the initial  coherence, a key contribution that was missed in the past \cite{scu03,scu11,har12,abe12,uzd15,kam16,cam19,fel03,pla14,kar16,bra16,bra17,bra20,rod19,san19,fra19,hor13,abe14,los15,nar15,los15a,cwi15,gou15,kor16,gou16,gou18,kwo18,los19,lob21}. We specifically identify an out-of-equilibrium regime 
where maximum work extraction is increased by  quantum coherence for fast processes beyond linear response,  and show that   the presence of coherence  is always detrimental  for strong 
thermalization. We finally illustrate our results with an analysis of a driven qubit.

%
%

{\emph{Dynamic Bayesian network.}} Let us consider a driven quantum system with time-dependent Hamiltonian $H_t= \sum_m \omega_m^t \ketbra{e_m^t}$, with instantaneous eigenvectors
$\ket{e_m^t}$ and  corresponding eigenvalues $\omega_m^t$. We assume that the initial populations  are  thermally distributed at inverse temperature $\beta$ in 
the energy basis  $\ket{e_m^{0}}$, but impose no restrictions on the off-diagonal elements, except the positivity of the state. The system may thus exhibit arbitrary quantum coherence in the energy basis. As a result, the initial system
density operator, $\rho_0 = \sum_i p_i^0 \ketbra{s^0_i}$, does not necessarily commute with the initial Hamiltonian, $[H_0,\rho_0] \neq 0$. The two eigenbases 
$\{\ket{s^0_i}\}$ and $ \{ \ket{e_m^0} \}$ are hence  not mutually orthogonal in general. This {makes} the analysis of the thermodynamics of the system in the energy eigenbasis nontrivial \cite{mic20,par20,str20,mic21}.  
We additionally suppose that the system is weakly coupled to a thermal reservoir,  at the same inverse temperature $\beta$, with density operator $\rho_R= \sum_\mu p_\mu \ketbra{e_\mu^R}$ and  Hamiltonian  $H_R$. We will use latin (greek) indices for system (bath) variables to distinguish the two. The interaction Hamiltonian $H_{SR}$ between system and reservoir is taken to satisfy strict energy conservation, $[H_t+H_R,H_{SR}]=0$, to ensure  weak coupling. We further denote by  $U_t$ the total  time evolution operator, $\partial_t U_t = - i (H_t + H_R + H_{SR})U_t$. The  instantaneous eigendecomposition of the system density operator at time $t$ then follows from the local evolution, $\rho_t = \Tr_R\{U_t (\rho_0 \otimes \rho_R) U_t^\dagger \}=\sum_j p_j^t \ketbra{s^t_j}$. 

In order to analyze the influence of quantum coherence on the nonequilibrium thermodynamics of the driven open  system, we next 
construct a  dynamic Bayesian network that describes the relationship between dynamical variables  through conditional probabilities evaluated via Bayes' rule \cite{nea03,dar09}. The conditional probability of initially finding  the system in the energy state $\ket*{e_m^0}$, given that it is in the eigenstate  $\ket*{{s^0_i}}$, is $ p(m^0|i^0) = \abs*{\braket*{e_m^0}{s^0_i}}^2$. Likewise, the conditional probability of finding  the system at a later time $t$ in the state $\ket*{e_n^t}$, given that it is in the eigenstate  $\ket*{{s^t_i}}$, reads  $p(n^t|j^t) = \abs*{\braket*{e_n^t}{s^t_j}}^2$. Since the reservoir is thermal, $\rho_R$ is diagonal in the energy basis,  implying that there exists a joint eigenbasis for the operators $\rho_R$ and $H_R$. The probability to initially find the bath in the eigenstate 
$\ket*{{e_\mu^R}}$ is accordingly $p_\mu$. The conditional probability for the  joint evolution of system and reservoir between time $0$ and $t$ then follows as $p(j^t,\nu|i^0,\mu) = \abs*{\mel*{s^t_j,e_\nu^R}{U_t}{s^0_i,e_\mu^R}}^2$.
For a given nonequilibrium driving protocol, we may now define a conditional trajectory $\Gamma = (s_i^0,s_j^t,e_m^0,e_n^t,e_\mu^R,e_\nu^R)$ for the composite system with path probability \cite{mic20,par20,str20,mic21}
\begin{align}\label{pf}
    P[\Gamma] = p_i^0 p_{\mu} p(m^0|i^0) p(j^t,\nu|i^0,\mu) p(n^t|j^t).
\end{align}
The above quantity contains the entire information about the quantum coherence of the system in the energy basis and its time evolution with the weakly coupled heat bath. We may also introduce a backward conditional trajectory $\Gamma^*=(s_j^t,s_i^0,e_n^t, e_m^0,e_\nu^R, e_\mu^R)$ by evolving the  state $\rho_t \otimes \rho_R$ with a 
time-reversed evolution, with path probability,
\begin{align}\label{pb}
   {P}^*[{\Gamma}^*] = p_j^t p_{\nu} p(n^t|j^t) p(i^0,\mu|j^t,\nu) p(m^0|i^0).
\end{align}
{\emph{Fluctuation relations with quantum coherence.}}
A detailed quantum fluctuation relation may   be derived by evaluating the ratio of forward and backward path probabilities, Eq.~\eqref{pf} and Eq.~\eqref{pb}, \cite{mic20,par20,str20,mic21}. We concretely find 
\begin{align}\label{ft}
    \frac{{P}[\Gamma]}{{{P^*}}[{\Gamma}^*]} = \frac{p^0_i p_\mu}{p^t_j p_\nu} = e^{\Delta s + {\Delta s_R}}=
     e^{ \beta (w - \Delta F) - \Delta c - d_t},
\end{align}
where the first equality follows from the microreversibility of the  unitary evolution of the composite system,  $p(i^0,\mu|j^t,\nu) = p(j^t,\nu|i^0,\mu)$. In the second equality, we have written the total stochastic entropy change of the composite system as the sum of the stochastic entropy variations of the system, $\Delta s = s_t-s_0=-\ln(p^t_j/p^0_i)$, and of  the reservoir, $\Delta s_R= -\ln(p_\nu/p_\mu)$ \cite{def11}. To obtain the third equality, we have introduced the stochastic heat exchanged with the bath,  $\beta q = \Delta s_R$, and formulated the system entropy, $ s_t = \beta (u_t - f_t)$, as a function of the stochastic internal energy $u_t = \mel*{s_t }{H_t}{s_t}$ and of the stochastic nonequilibrium free energy $f_t$. The nonequilibrium free energy may be further expressed as $\beta f_t = \beta F_t  +c_t + d_t$, where $F_t = - (1/\beta)\ln{Z_t}$ is the usual equilibrium free energy  and $c_t = \ln(p_i^t/{p_j^{t}}^{\text{d} }) $ is the stochastic relative entropy of coherence that {quantifies} the difference between the actual  state  $\rho_t$ and the associated diagonal state in the energy basis $ \rho_t^\text{d}$ \cite{str17}. The quantity $d_t = \ln({p_i^t}^{\text{d} } /{p_j^t}^{\text{eq}})$ is  furthermore the stochastic relative entropy that measures the lag between the nonequilibrium state $\rho_t^\text{d}$ and the corresponding equilibrium state $\rho_t^\text{eq}$ \cite{def11,kaw07,vai09}. After averaging over the forward process, the latter  reduce to familiar relative entropies,   
$\mathcal{C}_t = \avg{c_t} =   S(\rho_t||\rho^\text{d}_t)$ and $\mathcal{D}_t = \avg{d_t} =   S(\rho^\text{d}_t||\rho^\text{eq}_t)$
\cite{cov91}.

Equation \eqref{ft} is a quantum extension of the  detailed fluctuation theorem by Crooks \cite{cro99,tal09}, to which it reduces when forward and backward initial states are thermal, $c_0=c_t = d_t = 0$. The novel aspect of the relation \eqref{ft} is the inclusion of the difference, $\Delta c = c_t-c_0$, of  final and initial stochastic relative entropies of coherence, which was missed so far \cite{san19,fra19}. The presence of initial quantum coherence, quantified by $c_0$, strongly influences the work extraction properties of driven quantum systems, as we will discuss below. We note that the contribution of nonthermal initial populations may be easily added by replacing $d_t$ by $ \Delta d = d_t-d_0$. Integrating Eq.~\eqref{ft} over  all forward trajectories, we then obtain the integral quantum fluctuation relation,
\begin{align}\label{integra}
    \avg{e^{\beta(w - \Delta F) - \Delta c - \Delta d)}} = 1.
\end{align}
Expression \eqref{integra} is a fully quantum generalization of the Jarzynski equality \cite{jar97} for driven open quantum systems. It holds for arbitrary initial (and final) nonequilibrium states, with both nonthermal populations and quantum coherences in the energy basis. Its provides the foundation of our study of the energetics of quantum coherence.

{\emph{Quantum maximum-work theorem.}} Determining the maximum amount of work that a  system can deliver is a central task of  classical and quantum thermodynamics \cite{cal85}. Applying Jensen's inequality to Eq.~\eqref{integra}, we obtain
\begin{align}
\label{6}
    \beta(W - \Delta F) \geq \Delta \mathcal{C} + \Delta \mathcal{D},
\end{align}
where  $W = \avg{w}$ is the mean work---we use the convention that $W$  is positive when performed on the system. Equality is reached when the entropy production stemming from the difference between  the initial composite state $\rho_t \otimes \rho_R$ and the final state $U_t (\rho_0 \otimes \rho_R) U_t^\dagger$ vanishes \cite{man18}.
The maximum extractable work,  $-W$, is thus 
\begin{equation}
\label{7}
 \beta W_\text{max} = -\beta \Delta F - \Delta \mathcal{C} - \Delta \mathcal{D}= -\beta \Delta \mathcal{F},
 \end{equation}
 where we have defined the generalized quantum free energy $\mathcal{F} = F + kT(\mathcal{C}+ \mathcal{D})$ that extends the equilibrium free energy $F$ with contributions stemming from quantum coherence $\mathcal{C}$ and athermality $\mathcal{D}$ (with $\beta =1/kT$ and $k$ the Boltzmann constant); the latter quantity reduces to the  free energy introduced in Ref.~\cite{los15}, when  $\mathcal{D}=0$.
 We therefore obtain the general result that more work than the standard equilibrium work, $- \Delta F$, can only be gained from an arbitrary  quantum system  when the following (necessary) condition is satisfied:
 \begin{equation} 
\Delta \mathcal{C} + \Delta \mathcal{D}<0.
 \end{equation} For an initial thermal state, $ \mathcal{C}_0 = \mathcal{D}_0 = 0 $, we have $\Delta \mathcal{C} + \Delta \mathcal{D} =  \mathcal{C}_t +  \mathcal{D}_t\geq 0$ \cite{lan21}.  In other words, quantum coherence $\mathcal{C}_t$ induced, for example, through  the mechanism of quantum friction \cite{fel03,pla14},  and  athermality $\mathcal{D}_t$, generated during the time evolution \cite{spe13,alh16}, are both detrimental for quantum work production. One may hence conclude that only initial quantum coherence  $ \mathcal{C}_0 $ and initial athermality  $ \mathcal{D}_0$ are a potential resource for  work extraction in quantum thermodynamics. Expression \eqref{7} provides a quantum extension of the standard second law of thermodynamics \cite{cal85}.

%
%

{\emph{Unitary work extraction.}} We now analyze the conditions under which initial quantum coherence may be harnessed for useful work extraction. For simplicity, we first consider  the case of unitary dynamics by setting the system-bath coupling to zero, $H_{SR}=0$. We further assume that the initial populations are  thermal,  $\mathcal{D}_0=0$. We proceed with the  observation that adiabatic driving  leaves the density matrix elements of a system with  nondegenerate spectra unchanged in the instantaneous eigenbasis, except for a phase factor \cite{pin02}. The relative entropy of coherence remains accordingly constant, $\Delta \mathcal{C} = 0$, for an adiabatic transformation since it is phase independent. No useful work may  therefore be extracted from quantum coherence in this case. A general requirement for positive work extraction from initial quantum coherence is consequently that the unitary driving is nonadiabatic. The criterion for adiabatic dynamics, namely that the total evolution time (or duration of the driving protocol) $\tau_P$ ought to be much larger than the adiabatic time, defined as the timescale set by the square of the inverse gap, $\tau_A = \text{max}_{r\in[0,1]}|\langle e_m^r|\partial_r H_r|e_n^r\rangle |/|e_m^r-e_n^r|^2$, $\forall\, m\neq n$, with $r=t/\tau_P$ \cite{ami09,alb18}, should thus not be satisfied for coherence-enhanced work extraction. In other words, the driving time $\tau_P$ should be of the order of (or smaller than) the adiabatic time $\tau_A$:
\begin{equation} 
\label{9}
\tau_P  \lesssim \tau_A \quad \text{(unitary criterion). }
\end{equation}

\begin{figure}[t]
    \includegraphics[width=.48\textwidth]{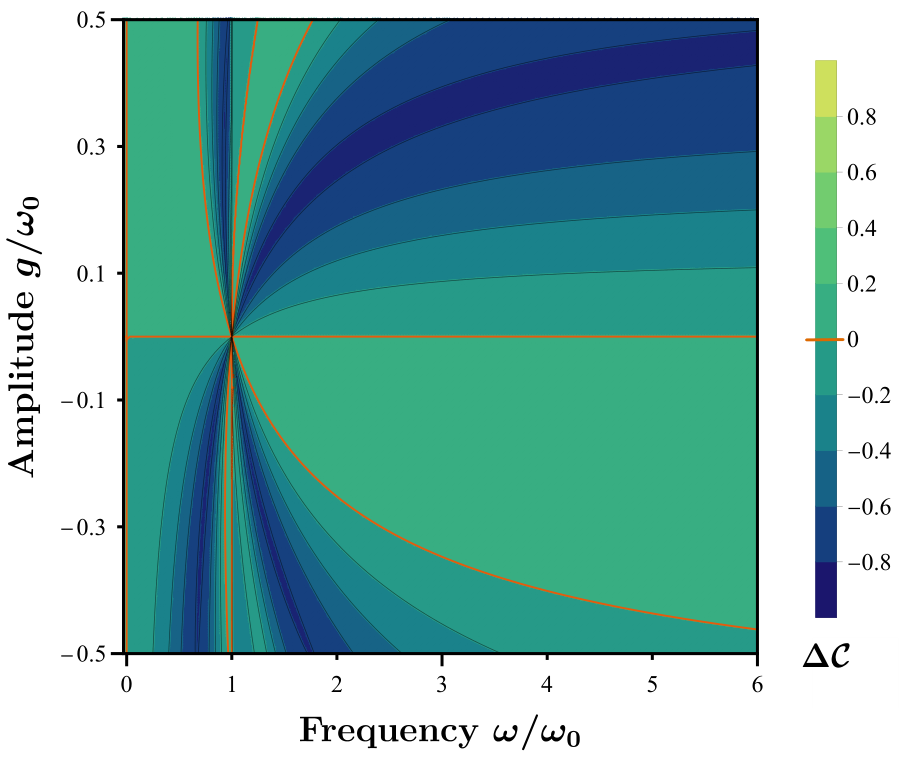}   
    \caption{Quantum coherence for a periodically driven qubit. Change of  relative entropy of coherence, $\Delta \mathcal{C}$, for a  two-level system in a rotating magnetic field as a function of driving frequency $\omega$ and  driving amplitude $g$ (with $\beta=0.5$): $\Delta \mathcal{C}=0$ along the orange lines (in particular, in the adiabatic limit $\omega\rightarrow 0$). For small driving amplitude $g$, $\Delta \mathcal{C}<0$ close to resonance $\omega\simeq\omega_0$, enabling coherence-to-work conversion.}
    \label{fig:unitary}
  \end{figure}
  
We illustrate the above discussion with the example of a spin-1/2  in a rotating magnetic field with Hamiltonian $H_t= (\omega_0/{2})\sigma_z + ({g}/{2}) [\cos (\omega t) \sigma_x + \sin (\omega t) \sigma_y]$, where $\omega_0$ is the frequency 
of the two-level system, $\omega$ and $g$ are the respective frequency and amplitude of the driving field, and $\sigma_{x,y,z}$ are the usual Pauli operators \cite{oli07}. The duration of the driving protocol is taken to be $\tau_P =2 \pi/\omega$. We choose an initial state, $\rho_0 = \rho_\text{th} + \chi$, that is thermal in the initial energy basis, plus a nondiagonal matrix $\chi$ of elements $a \sqrt{p_\text{ground}(1-p_\text{ground})}$, for $a \in [0,1]$ ranging from incoherent to maximally coherent. The change of relative entropy of coherence $\Delta \mathcal{C}$ at half the Rabi frequency $\Omega = \sqrt{g^2 + (\omega_0-\omega)^2}$ is shown as a function of the driving frequency and of the driving amplitude in Fig.~\ref{fig:unitary}. As expected,  $\Delta \mathcal{C} = 0$ for adiabatic driving $\omega \rightarrow 0$ (or, equivalently,  $\tau_P \rightarrow \infty$) (vertical orange line on the left). We moreover note that, for small driving amplitudes, $\Delta \mathcal{C} < 0$ occurs around resonance,  $\omega \simeq \omega_0$, and that $\Delta \mathcal{C}$ typically decreases for increasing $|\omega-\omega_0|$ and $|g|$. These are the areas where   quantum coherence may be converted into  work. For a given amplitude $g$, maximum work extraction is concretely achieved for the driving frequency (Supplemental Material)
\begin{equation}
 \omega_\text{opt} = \frac{E^2\left( \omega_0 + g e^{-\beta E/2} \right)}{E^2 - 2g\left( \omega_0 \sinh{({\beta E}/{2})} + g\right)},
 \end{equation} with the energy $E= \sqrt{g^2 + \omega_0^2}$. The optimal frequency  $ \omega_\text{opt}\simeq g \cosh(\beta \omega/2)$ scales linearly with $g$ for small $g$.
 
 \begin{figure*}[t]
	\centering
	\begin{tikzpicture}
	\node (a) [label={[label distance=-.4 cm]145: \textbf{a)}}] at (0,0) {\includegraphics[width=0.32\textwidth]{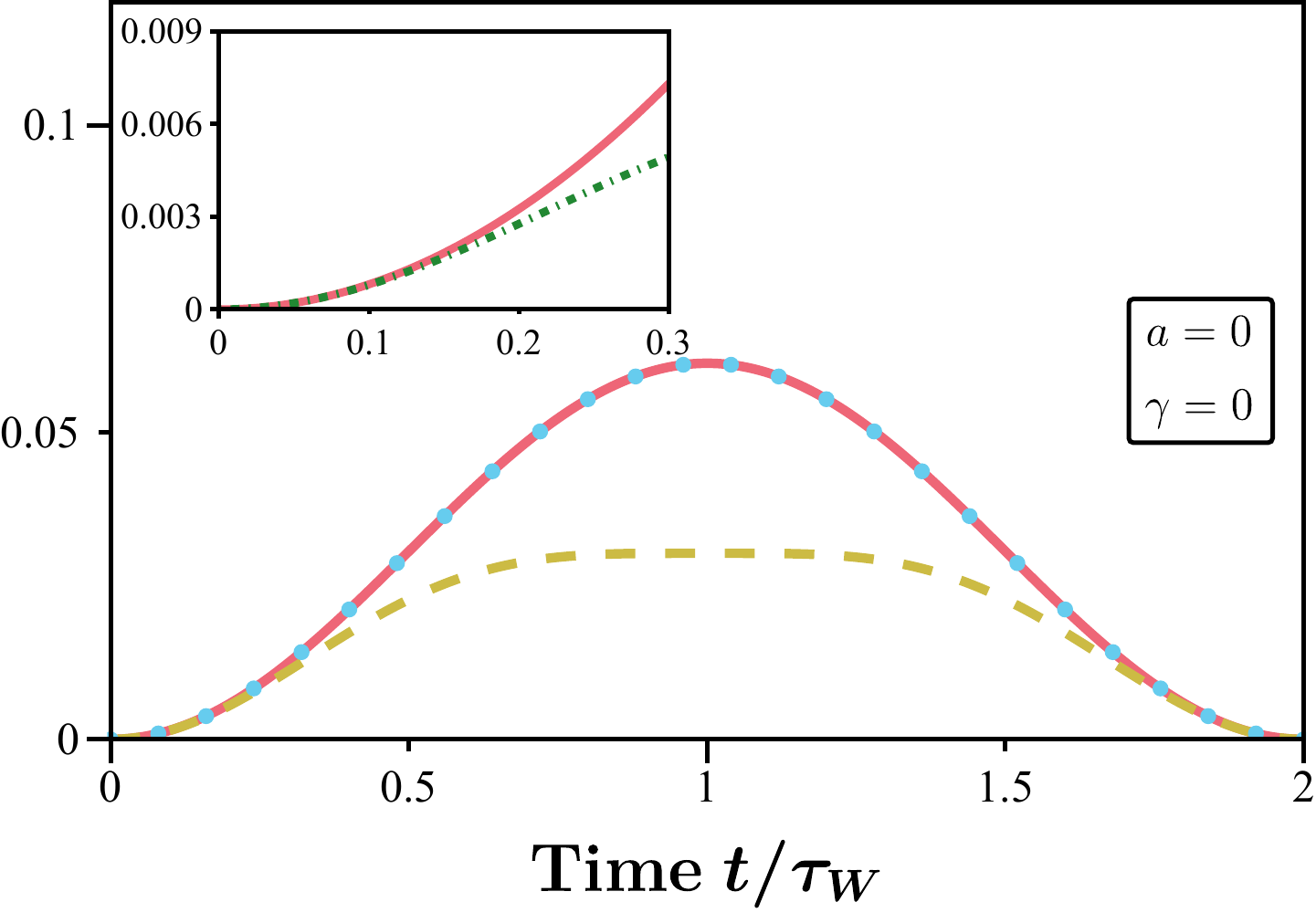}};	\node (a) [label={[label distance=-.4 cm]145: \textbf{b)}}] at (6,0) {\includegraphics[width=0.32\textwidth]{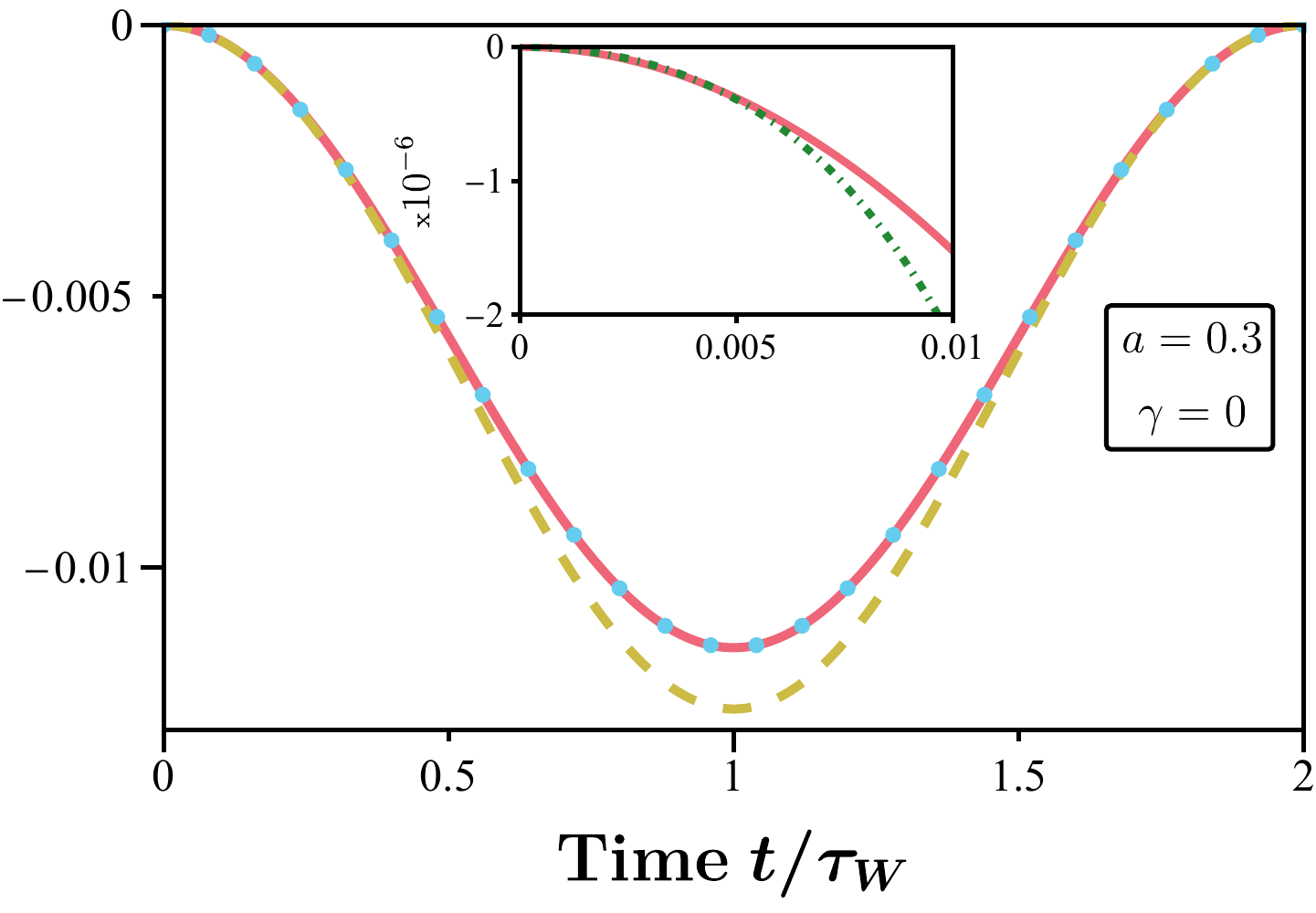}};
	\node (a) [label={[label distance=-.4 cm]145: \textbf{c)}}] at (12.0,0) {\includegraphics[width=0.32\textwidth]{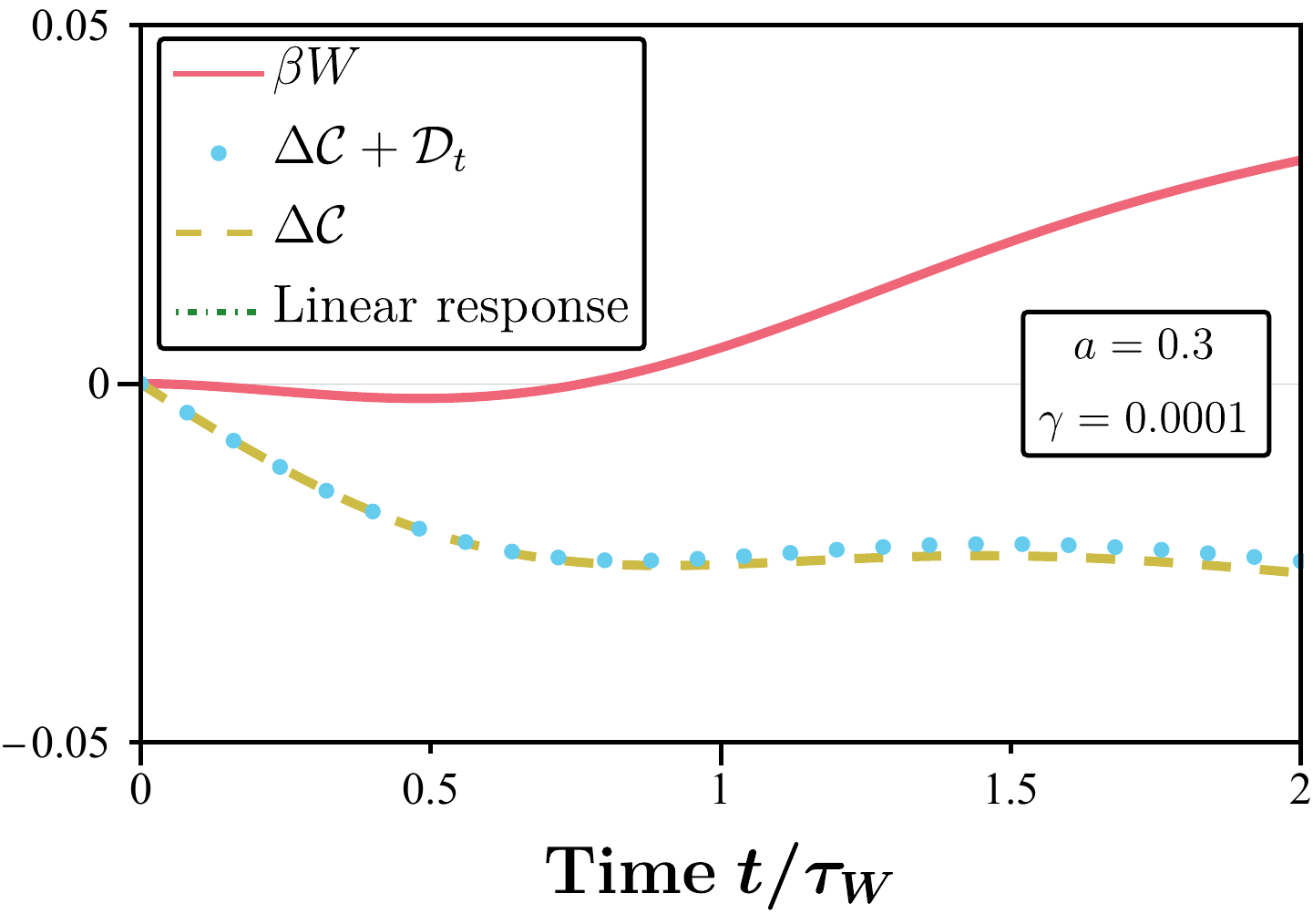}};
	\end{tikzpicture}
	\caption{Coherence-to-work conversion for a periodically driven qubit. a) Without initial coherence ($a=0$), work is consumed, $\beta W>0$, as quantum coherence is created, $\Delta \mathcal{C}>0$, during unitary evolution. b) With initial coherence ($a=0.3$), work is efficiently extracted from quantum coherence, $\beta W<0$ and $\Delta \mathcal{C}<0$. Maximum work production occurs at half the Rabi time, $\tau_W=\Omega/\pi$, where  $\Omega$ is the Rabi frequency. This time lies beyond the linear response regime; insets show deviations from the quantum fluctuation-dissipation relation for work. c) For nonunitary dynamics ($\gamma \neq0$), coherence-to-work is hampered by decoherence, which reduces $\Delta \mathcal{C}$, and by nonequilibrium entropy production, which leads $\beta W$ to deviate from  the variation, $\Delta \mathcal{C} +  \mathcal{D}_t$, of coherence and athermality.
    Parameters are $\omega = 1$,  $g = 0.005$, $\beta = 0.5$ and $\delta = \omega_0-\omega=- 0.005$.}\label{fig2}
\end{figure*}
 
In order to gain additional insight, we  display in  Fig.~2  the time evolution of the average work, $\beta W$ (red), the change of quantum coherence, $\Delta \mathcal{C}$ (yellow), and the added variations  of coherence and athermality,  $\Delta \mathcal{C} +  \mathcal{D}_t$ (blue); $\Delta F=0$ for the periodic driving considered. In the absence of initial quantum coherence ($a=0$),  $\beta W >0$ and  no  work can  hence be gained from the system (Fig.~2a). In this scenario, work is consumed to create coherence, $\Delta \mathcal{C}>0$. By contrast, for  $a=0.3$,   quantum coherence is successfully converted into mechanical   work,  $\beta W < 0$ with $\Delta \mathcal{C}<0$   (Fig.~2b). We mention that inequality \eqref{6} is here saturated. The upper bound for the maximum work \eqref{7} is therefore reached.  In general, nonequilibrium    entropy production, associated with the athermality $\mathcal{D}_t$ of the system, reduces the efficiency of the coherence-to-work conversion (seen as the difference between red and yellow lines in Fig.~2b).

An important observation is that maximum work extraction occurs at a time $\tau_{W}$ (here given by half the Rabi time, $ \tau_{R}/2= \pi/\Omega$), which lies beyond the linear response regime (Fig.~2b): In the absence of initial coherence ($a=0$), the  linear response approximation leads to the fluctuation-dissipation relation $W = \Delta F +\beta \sigma^2_W /2 - {\cal Q}_0$, where $\sigma^2_W$ denotes the work variance and ${\cal Q}_0= (\beta/2) \int_0^1 dy I^y(\rho_t^\text{eq},\Delta  H_t)$, is a non-negative quantum correction than only vanishes when $[H_t,\dot H_t]=0$ \cite{mil19,sca20}; the Wigner-Yanase skew information quantifying the quantum uncertainty of observable $L$, measured in  state $\rho$, is here given by $I^y(\rho,L)= \tr{[\rho^y,L][\rho^{1-y},L]}$ \cite{wig63}. For $a\neq 0$, the fluctuation-dissipation relation is generalized to  $W = \Delta F +\beta \sigma^2_W /2 - {\cal Q}_0 - E_{\cal Q}$, with  an additional contribution, $E_{\cal Q} = \tr{\Delta H_H \chi}$, stemming from the initial coherence, where $\Delta H_H$ is the variation of the Hamiltonian in the Heisenberg picture \cite{rod22}. In both cases, the linear response approximation only agrees with the exact work for $t\ll \tau_W$ (insets of Figs.~2ab).

%
%

{\emph{Nonunitary work extraction.}}
The problem becomes more complicated when the system is coupled to a heat bath. In this situation, the available amount of quantum coherence is  suppressed by environment-induced decoherence \cite{zur03,sch07}, and $\Delta \mathcal{C}$ is reduced compared to the unitary evolution (yellow line in Fig.~2c). Entropy dissipation associated with system-bath correlations, as well as  entropy production caused by the athermality of the reservoir \cite{pta19} further hinder coherence-to-work conversion (large difference between red and yellow lines in Fig.~2c). The unitary criterion \eqref{9} is thus not enough to guarantee  successful coherence-to-work transfer in this case. Under Markovian dynamics, the decoherence time scale is  longer than the bath relaxation  time. The decoherence rate may accordingly be expressed for short times as $1/\tau_D =  -2 \text{tr}[\rho_0 \dot \rho_0]/\text{tr}[\rho_0^2]$ \cite{xu19} (see also Refs.~\cite{kim96,tol04,bea17,gu17}), and the contribution from bath athermality  may be neglected. As a consequence, coherence-to-work conversion in open quantum systems with nonunitary dynamics  is only  effective when the work extraction time $\tau_W$ is much shorter than the decoherence time scale $\tau_D$:
\begin{equation}
\label{10}
\tau_W\ll \tau_D \quad \text{(nonunitary criterion).}
\end{equation}

We illustrate the predictive power of condition \eqref{10} by taking the same driven two-level example as before and letting it weakly interact, with coupling strength $\gamma$, to a bath of infinitely many harmonic oscillators at inverse temperature $\beta$ \cite{tan98}. Considering that the relaxation time of the reservoir is short  
compared with the timescale of the system in the rotating frame, specified through the unitary operator $U_r = \exp {-i \omega t \sigma_z/2}$, the master equation for  
$\tilde{\rho}_t = U_r \rho_t U_r^\dagger$ is of the standard Markovian form, $\dot{\tilde{\rho}}_t = - i [\tilde{H}_{\text{eff}},\tilde{\rho}_t] + \mathcal{L}[\tilde \rho_t]$, with the  dissipator, $\mathcal{L}[\tilde \rho_t] =\gamma \bar{n} D_+[\tilde{\rho}_t] + \gamma (\bar{n} + 1) D_-[\tilde{\rho}_t]$,  
where $\tilde{H}_{\text{eff}} = U_r H_t U_r^\dagger - \omega \sigma_z/2$ is the effective system Hamiltonian in the rotating frame and 
$D_\pm[\mathcal{O}] = \sigma_\pm \mathcal{O} \sigma_\mp -\{ \sigma_\mp \sigma_\pm , \mathcal{O} \}/2$ are the dissipative channels for the $\sigma_\pm$ transitions ($\bar{n}$ denotes the mean number of bath excitations) \cite{tan98}. The average work flow of the system may then be consistently defined via the first law as $\dot{W}_t = \dot{E}_t - \dot{Q}_t$, where $E_t = \tr{H_t\rho_t}$ is the internal energy   and $\dot{Q}_t = \tr{H_t \mathcal{L}[\rho_t]}$ is the heat flow \cite{elo20}.

Figure 3 depicts the average work $\beta W$ as a function of time for three values of the coupling  strength. For  small damping,  $\tau_D = 5 \tau_W$ (blue), coherence is efficiently converted into useful work, $\beta W <0$, and maximal work extraction occurs at time $\tau_W$ like in the unitary regime shown in Fig.~2b. For moderate  damping, $\tau_D = \tau_W$ (green) and $\tau_D = 0.5 \tau_W$ (red),  a strongly diminished amount of work can be produced at short times owing to the adverse effect of decoherence (the decoherence time $\tau_D$ is represented by the vertical dashed lines) \cite{com}.  Finally, for strong thermalization, $\tau_D \ll \tau_W$ (not shown), no work can be effectively extracted from the system, and quantum coherence is always detrimental as in Fig.~2a.

\begin{figure}[t!]
   \includegraphics[width=.475\textwidth]{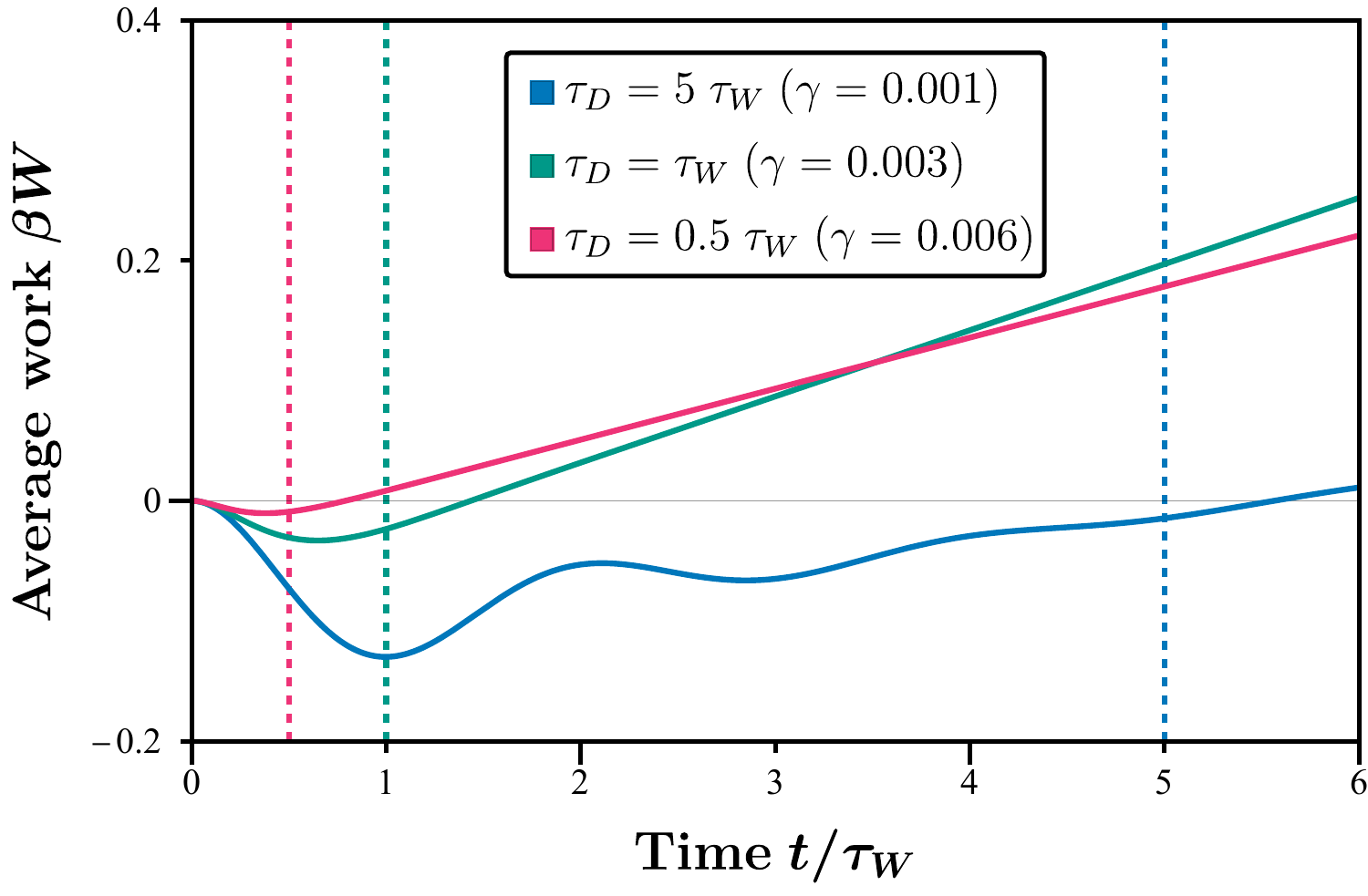}%
    \caption{Influence of decoherence on  work extraction. Coherence-to-work conversion, $\beta  W<0$, is only effective when the work extraction time $\tau_W$ is much smaller than the decoherence timescale $\tau_D$, $\tau_W\ll \tau_D$.  No work can be extracted for strong thermalization, $\tau_D\ll\tau_W$ (vertical lines indicate the decoherence time $\tau_D$). Same parameters as Fig.~2.}
    \label{fig:timescales}
    \end{figure}

%
%
\textit{Conclusions.} 
We have performed a detailed investigation of the interconversion of quantum coherence and mechanical work  in nonequilibrium quantum  processes, and examined the conditions under which quantum coherence is a useful resource in quantum thermodynamics.  We have, in particular, derived a novel maximum-work theorem from a generalized fluctuation relation, and obtained explicit criteria for successful coherence-to-work conversion, for both closed and open quantum systems. Our results highlight the competing influence of initial coherence (that can be converted into work) and coherence generated during time evolution through quantum friction (that consumes work), as well as the adverse effects of entropy production and decoherence. We have additionally discerned a timescale for optimal coherence-enhanced work extraction that lies beyond the range of linear-response regime. These findings emphasize the importance of initial quantum coherence for thermodynamic applications and their generation via reservoir-engineering techniques \cite{mya00,kra11,mur12,sha13,lin13,har22}.
 Such coherent (and, possibly, athermal) baths may be easily described in the dynamic Bayesian network formalism by including a contribution $\Delta c_ \text{bath} +  \Delta d_ \text{bath}$ in the fluctuation relations. We expect these insights to be useful for the  design of efficient quantum-enhanced nanomachines.


\begin{acknowledgments}
    We acknowledge  financial support from the German Science Foundation (DFG) (under project FOR 2724) and thank Kaonan Micadei for useful discussions.
 \end{acknowledgments}


\pagebreak
\widetext

\newpage 
\begin{center}
\vskip0.5cm
{\large \bf Supplemental Material: Nonequilibrium thermodynamics of quantum coherence beyond linear response}
\end{center}


The Supplemental Material provides details about the analytical solution and the quantum thermodynamics of the driven two-level system, as well as a discussion of the decoherence timescale.\\

\section{I. Driven two-level system} 

We consider a two-level system, with frequency $\omega_0$,  in a rotating magnetic with  Hamilton operator
\begin{align}
    H_t = \frac{\omega_0}{2}\sigma_z + \frac{g}{2} \left[\cos (\omega t) \sigma_x + \sin (\omega t) \sigma_y\right],
\end{align}
where $\omega$ is the driving frequency and $g$ the driving amplitude \cite{oli07}. The solution to its time evolution operator $\partial_t U_t = - i H_t U_t$ is explicitly given by
\begin{align}
    U_t = \exp{-i \frac{\omega t}{2}\sigma_z}\exp{-\frac{i}{2}(\delta \sigma_z + g \sigma_x)t},
        \label{first rotation}
\end{align} 
where $\delta = \omega_0-\omega$ is the detuning between the driving frequency and the natural frequency of the two-level system. 
We define for future reference the transformation to the instantaneous energy basis of $H_t$
\begin{equation} R^H_t = 
    \left( \begin{matrix}
    \cos{\frac{\theta}{2}} & \sin{\frac{\theta}{2}}e^{-i \omega t} \\
    -\sin{\frac{\theta}{2}}e^{i \omega t} & \cos{\frac{\theta}{2}}, 
    \end{matrix}  \right)
\end{equation}
with the angle $\theta = \arctan{g/\omega_0}$. We take an initial 
 state $\rho_0$ whose populations are thermally distributed at inverse temperature $\beta$, and with coherences $a\in[0,1]$. It is convenient to write this state in the form
   \begin{align}
\rho_0= \frac{1}{2} \left( 1 - \tanh{\frac{\beta E}{2}} \sigma'_z + a \sech{\frac{\beta E}{2}}  \sigma'_x\right),
\label{thc}
\end{align}
where $\sigma'_i = R^{H\dagger}_0 \sigma_i R^H_0$ are the Pauli matrices rotated to the basis of the initial Hamiltonian and $E = \sqrt{g^2+\omega_0^2}$
is the initial energy gap. 
The term $a$ gauges the amount of coherences of the state: if $a = 0$, $\rho_{0}$ reduces to a standard thermal state, whereas, if $a=1$, $\rho_{0}$ is 
pure. We choose real  coherences for simplicity---any type of coherence may be  included via by a $R^z_\phi = e^{-i \phi \sigma_z/2}$ rotation on $\sigma_x$, 
without providing additional physical insight.

It is advantageous to analyze the dynamics in the rotating frame. Any operator $\mathcal{O}$ becomes accordingly 
$\tilde{\mathcal{O}} = R^{z\dagger}_{\omega t} R^H_t \mathcal{O} R^{H\dagger}_t R^{z}_{\omega t}$. Combining Eqs.~\eqref{first rotation} and \eqref{thc}, we obtain the density operator
\begin{align}
\label{state}
    \tilde{\rho}_t = \frac{1}{2} \left[1 - \tanh{\frac{\beta E}{2}} m_t^z + a \sech{\frac{\beta E}{2}}  m_t^x\right],
\end{align}
where the operators $m_z$ and $m_x$ are given by 
\begin{align}
    m_t^z &= \sigma_z - 2\mu_t N_t,\\
    m_t^x &= \sigma_x - 2\nu_t N_t,
\end{align}
We have here defined the following quantities
\begin{align}
    N_t &= \mu_t \sigma_z - \frac{\delta}{\abs{\delta}} \sqrt{1-\mu^2_t} \,R_z^\dagger(\xi_t) \sigma_x R_z(\xi_t),\\
    \mu_t &= \frac{g \omega}{E \Omega} \sin (\frac{\Omega t}{2}),\\
    \nu_t &= -\frac{E^2 + \Omega^2-\omega^2} {2 E \Omega} \sin (\frac{\Omega t}{2}),\\
    \xi_t &= \arctan \frac{2g \omega \cot (\frac{\Omega t}{2})}{E^2+\Omega^2-\omega^2}.
\end{align}
We emphasize that, in this local Hamiltonian basis, all coherences are energetic by construction, simplifying analytical calculations. We further note that $N_t$ disappears in the adiabatic limit ($\omega \rightarrow 0$). 
In this scenario, the initial state undergoes a simple phase shift in the local Hamiltonian basis.

We next evaluate the  average work performed on the system during time $t$, $W = \tr{\tilde{\rho}_t \tilde{H}}-\tr{\tilde{\rho}_0 \tilde{H}}$, where $\tilde{H} = {E}\sigma_z/2$. Using Eq.~\eqref{state}, we find:
\begin{align}\label{example:work}
    W_t = \left( \tanh \frac{\beta E}{2} + a \frac{E^2+\Omega^2-\omega^2}{2 g \omega} \sech \frac{\beta E}{2}
     \right) \frac{g^2 \omega^2}{E^2 \Omega^2} \sin^2 \frac{\Omega t}{2}.
\end{align}
We observe that $W \rightarrow 0$ in the adiabatic limit $\omega \rightarrow 0$, as discussed in the main text for general systems.
The  condition for work extraction, $W<0$, furthermore yields  
\begin{align}
   \sinh \frac{\beta E}{2} < \frac{\omega^2-(E^2+\Omega^2)}{2 g \omega} \; a.
\end{align}
We recover the impossibility of extracting useful work from an incoherent thermal state, since the inequality cannot be fulfilled for  $a = 0$ and ${\beta E}>0$.

By minimizing Eq.~\eqref{example:work} with respect to the driving frequency $\omega$, we may additionally derive an expression for the optimal driving frequency that allows for maximum work production from initial quantum coherence for a given driving amplitude $g$. We obtain 
\begin{align}
    \omega_\text{opt} = \frac{E^2\left( \omega_0 + g e^{-\frac{\beta E}{2}} \right)}{E^2 - 2g\left( \omega_0 \sinh{\frac{\beta E}{2}} + g\right)},
\end{align}
in which case the state at the end of half Rabi period is the ground state of the final Hamiltonian.

Finally, we evaluate  the time average of the work $W$, Eq.~\eqref{example:work}, both over the duration of the driving protocol $\tau_P$
and over the Rabi time $\tau_R$:
\begin{align}
    \bar{W}_R &= \frac{1}{\tau_R}\int_0^{\tau_R} \dd{t} W_t = \frac{g^2 \omega^2}{2 E^2 \Omega^2} \left( \tanh \frac{\beta E}{2} + a \frac{E^2+\Omega^2-\omega^2}{2 g \omega} \sech \frac{\beta E}{2}
    \right),\\
    \bar{W}_P &= \frac{1}{\tau_P}\int_0^{\tau_P} \dd{t} W_t = \left( 1 - \text{sinc} \; \frac{2\pi \Omega}{\omega} \right) \bar{W}_R,
\end{align}
where $\text{sinc} \, x = \sin{x}/x$. The two averages  retain the negativity condition of the previous discussion. However, for small 
values of $2\pi \Omega/\omega$, $\bar{W}_P$ becomes increasingly smaller, given the quadratic behavior of $\text{sinc} \, x$ near the origin. 
This might be specially important in the engineering of heat engines that make use of these initial coherences, where the average is 
usually performed with respect to the driving frequency \cite{bra16,bra17}.

\section{II. Decoherence timescale for the driven two-level system}
In this section, we compute  the decoherence timescale, $1/\tau_D =  -2 \text{tr}[\rho_0 \dot \rho_0]/\text{tr}[\rho_0^2]$ \cite{xu19}, 
for the driven two-level system weakly coupled to a Markovian reservoir. We first write  
\begin{align}\label{supp:decoherence}
    \tau_D = \frac{\text{tr}[\rho_0^2]}{2 \sum_i \gamma_i \widetilde{\text{cov}}_{\rho_0}\left( L_i^\dagger, L_i\right)},
\end{align}
where $\rho_0$ is the initial state of the evolution, $\gamma_i$ are the decay rates associated with the $L_i$ Lindblad operators in the
master equation, and $\widetilde{\text{cov}}_{\rho_0}\left( X, Y\right) = \text{tr}[\rho_0 X Y \rho_0] - \text{tr}[X \rho_0 Y \rho_0]$ is
 a generalized covariance \cite{xu19}.

For the damped two-level system, we respectively have $L_i=\{\sigma_-,\sigma_+\}$ and $\gamma_i=\{ \gamma(\bar{n}+1),\gamma\bar{n}\}$
where $\bar{n} = (\exp (\beta \omega_0) - 1)^{-1}$ is the thermal mean occupation number  at frequency $\omega_0$ and inverse temperature $\beta$.
We further express the  initial state as $\rho_0 = (1 + \textbf{r} \cdot \bm{\sigma})/2$, where $\textbf{r} = (r_\perp,r_z)$ is the Bloch vector
and $\bm{\sigma} = (\sigma_\perp,\sigma_z)$; the subscript $\perp$  refers to all perpendicular contributions  to $\sigma_z$.
Combining all the terms in Eq.~\eqref{supp:decoherence}, we find
\begin{align}
    1/\tau_D = \frac{4 \gamma}{1+r^2} \left[ (\bar{n}+1) \widetilde{\text{cov}}_{\rho_0}\left( \sigma_+, \sigma_-\right) + 
    \bar{n} \widetilde{\text{cov}}_{\rho_0}\left( \sigma_-, \sigma_+\right) \right] \quad \text{with}\quad 
    \widetilde{\text{cov}}_{\rho_0}\left( \sigma_\pm, \sigma_\mp\right) = \left( \frac{1 \mp r_z}{2}\right)^2 - \frac{1-r^2}{4},
\end{align}
where $r = \abs{\textbf{r}}$. Further simplifications lead to
\begin{align}\label{midway_decoherence}
    1/\tau_D = \frac{2}{1+r^2} \left[ \gamma (\bar{n}+1/2)(r^2 + r_z^2) + \gamma r_z \right].
\end{align}

Assuming that  the populations of the two-level system are thermal with respect to a bath of mean occupation number $\bar{m}$, then
$r_z = -1/(2\bar{m}+1)$ and we can rewrite Eq.~\eqref{midway_decoherence} as
\begin{align}
    1/\tau_D = \frac{\gamma (\bar{n}+1/2)}{\mathcal{P}} \left[ r_\perp^2 - 
    2\frac{1}{\bar{m}-1/2}\left(\frac{1}{\bar{n}-1/2}-\frac{1}{\bar{m}-1/2}\right) \right],
\end{align}
where $\mathcal{P} = (1+r^2)/2$ is the purity of the initial state. The first term in the square bracket is the contribution to the decoherence
time from the actual coherence of the state, while the second term is the contribution from the mismatch of the thermal occupations between
system and environment. We mention that the prefactor $\gamma (\bar{n}+1/2)$ is the usual decoherence time
considered in the optical Bloch equations \cite{tan98}. In fact, we recover it in the case where the initial state is maximally 
coherent in the energy basis, thus $r_z = 0$ and $r_\perp = r = 1$ in Eq.~\eqref{midway_decoherence}.


\begin{thebibliography}{99}
\bibitem{com} For $t\geq\tau_D$, the average work increases linearly in time, since the  system reaches a steady state and the entropy production rate is accordingly constant.
\end{thebibliography}
\end{document}